\documentclass{IEEEtran}
\usepackage{cite}
\usepackage{xcolor}
\usepackage{cite}
\usepackage{amsmath,amssymb,amsfonts}
\usepackage{algorithmic}
\usepackage{graphicx}
\usepackage{textcomp}
\usepackage{scalerel}
\usepackage{subfigure}
\usepackage{enumerate}
\usepackage{mathtools}
\usepackage{amsmath}

\def\BibTeX{{\rm B\kern-.05em{\sc i\kern-.025em b}\kern-.08em
    T\kern-.1667em\lower.7ex\hbox{E}\kern-.125emX}}
\begin{document}
\title{Omni-directional Pathloss Measurement Based on Virtual Antenna Array with Directional Antennas}
\author{Mengting Li, Fengchun Zhang, Xiang Zhang, Yejian Lyu and Wei Fan
\thanks{Mengting Li, Fengchun Zhang, Yejian Lyu and Wei
Fan are with the Antenna Propagation and Millimeter-wave Systems (APMS)
section, Aalborg University, Denmark.}
\thanks{Xiang Zhang is with China Academy of Information and Telecommunications
Technology (CAICT), Beijing 100191, China}
\thanks{Corresponding author: Wei Fan (Email: wfa@es.aau.dk).This work has been submitted to the IEEE for possible publication. Copyright may be transferred without notice, after which this version may no longer be accessible.}}

\maketitle

\begin{abstract}
Omni-directional pathloss, which refers to the pathloss when omni-directional antennas are used at the link ends, are essential for system design and evaluation. In the millimeter-wave (mm-Wave) and beyond bands, high gain directional antennas are widely used for channel measurements due to the significant signal attenuation. Conventional methods for omni-directional pathloss estimation are based on directional scanning sounding (DSS) system, i.e., a single directional antenna placed at the center of a rotator capturing signals from different rotation angles. The omni-directional pathloss is obtained by either summing up all the powers above the noise level or just summing up the powers of detected propagation paths. However, both methods are problematic with relatively wide main beams and high side-lobes provided by the directional antennas. In this letter, directional antenna based virtual antenna array (VAA) system is implemented for omni-directional pathloss estimation. The VAA scheme uses the same measurement system as the DSS, yet it offers high angular resolution (i.e. narrow main beam) and low side-lobes, which is essential for achieving accurate multipath detection in the power anglular delay profiles (PADPs) and thereby obtaining accurate omni-directional pathloss. A measurement campaign was designed and conducted in an indoor corridor at 28-30 GHz to verify the effectiveness of the proposed method.      
\end{abstract}


\section{Introduction}
\label{sec:introduction}
\IEEEPARstart{M}{illimeter} wave (mmWave) and sub-Terahertz (THz) frequency bands will continue to play a vital role in the future communication systems since they can provide vast frequency resources to enable high data rate transmission \cite{rappaport2013millimeter,degli2014ray}. To achieve optimal system deployment, omni-directional pathloss models which offer flexibility to superimpose arbitrary antenna patterns on the desired propagation scenarios are indispensible for system simulation. 
However, the omni-directional antennas, especially for those with horizontal polarization, cannot be easily obtained in mmWave and sub-THz bands due to the high design complexity and constrained fabrication accuracy. Besides, omni-directional antennas provide limited link budget in the measurements due to the low antenna gain. This issue becomes more pronounced in mmWave and sub-THz bands owning to the high pathloss. 
Therefore, the prevalent method to achieve omni-directional pathloss in high frequency bands is to synthesize the omni-pathloss based on directional channel sounding (DSS) \cite{gustafson2013mm,ling2017double,he2019propagation,kyro2011measurement,akdeniz2014millimeter,sun2015synthesizing,maccartney2015indoor,fuschini2017analysis,hur2014synchronous,hur2015wideband } i.e., with a directional antenna placed on a mechanical rotator to capture the signals from different directions. 

There are different ways to construct omni-directional pathloss based on DSS. In \cite{kyro2011measurement,akdeniz2014millimeter,sun2015synthesizing,maccartney2015indoor,fuschini2017analysis}, the received power of the omni-directional antenna is synthesized by summing up all the power spectrum over delay and angular domain above the noise level. This method is straightforward, but the same paths will be repeatedly counted during the rotation process resulting in underestimation of the omni-directional pathloss due to the wide main beam. In \cite{hur2014synchronous,hur2015wideband}, discrete propagation paths are extracted based on the measured results and used to calculate the estimated pathloss by summing up the powers of the identified paths. However, the path identification becomes problematic in angular domain with the non-ideal antenna pattern embeded in the measured frequency response. The omni-directional pathloss is synthesized by first identifying the paths in delay domain and then finding the maximum power at each identified delay bin in power angular delay profile (PADP). With limited delay resolution, the paths with similar delays cannot be totally detected and the omni-directional pathloss might be overestimated.  
In \cite{haneda2016estimating}, the power of diffuse paths are considered when approximating the omni-directional pathloss. However, it is still relying on accurate detection of propagation paths at the first step, which cannot be fulfilled by DSS system. 

In this paper, the omni-directional pathloss is estimated based on a virtual antenna array (VAA) concept with directional antennas instead of the conventional DSS scheme. With high angular resolution and low side-lobes offered by the VAA scheme, the multipath components can be recognized by finding local maxima on PADP and the omni-directional pathloss can be then estimated from the identified paths. The measurement campaigns are described in Section \ref{sec:campaign}. The theory of synthesizing the omni-directional path loss is given in Section \ref{sec:theory}. The results of estimated omni-directional pathloss using the proposed method based on VAA scheme and two conventional methods based on DSS scheme are compared and discussed in \ref{sec:results}. Finally, a conclusion is drawn in Section \ref{sec:conclusion}.

\section{Omni-directional Pathloss Synthesizing Theory}
\label{sec:theory}
Assume that a uniform circular array (UCA) consists of $P$ directional antenna elements which are uniformly distribute along a circle of radius $r$ on the $xoy$ plane. The center of the UCA is located at the origin of the coordinate system. The angular position of the $p$-th element is $\varphi_p = 2\pi \cdot (p-1)/P, p\in[1,P].$ 
The frequency band is within the range of [$f_L$, $f_U$] with $f_L$ and $f_U$ denoting the lower and the upper frequencies, respectively. Suppose there are $K$ plane waves impinging on the UCA with the $k$-th path from direction $\phi_k$. The multipath in our measurements given in Section \ref{sec:campaign} is mainly confined in the azimuth plane since the transmit and receive antennas are placed in the same height and the receive antenna has narrow beamwidth in the E-plane. Therefore, the discussion in this letter is limited to a 2-D model, i.e. with the multipath and the UCA confined in the UCA plane ($xoy$ plane herein). Note that it is possible to extend the present model into a 3-D model. The channel frequency response (CFR) at the $p$-th UCA element can be expressed as
\begin{equation}
\label{eq:1}
H_{p}(f) = \sum_{k=1}^{K} \alpha_{k} \exp(-j2 \pi f \tau_{k})\ \cdot a_{p}(f,\phi_k),
\end{equation}
where $\alpha_k$ and $\tau_k$ represent the complex amplitude and delay of the $k$-th path ($k\in [1,K]$), respectively. $a_p(f,\phi_k)$ is the transfer function between the $k$-th path and the $p$-th array element, which is normalized by that between the $k$-th path and UCA center, expressed as

\begin{equation}
\label{eq:2}
 a_p(f,\phi_k)= \exp( \frac{j2 \pi fr }{c}cos(\phi_{k}-\varphi_p)) g(f, \phi_k-\varphi_p),
\end{equation}
where $c$ is the speed of light. $g(f, \phi)$ is the 2-D complex radiation pattern (antenna gain is normalized to 1 for the ease of calculation) of the antenna element at $xoy$ plane with its boresight direction along $x$ axis. 

Using the modified classical beamforming proposed in \cite{li2022virtual}, the steering weight of the $p$-th array element can be written as
\begin{equation}
\label{eq:3}
\omega_p (f, \phi) = \exp[- \frac{j2 \pi fr }{c}cos(\phi-\varphi_{p})]\cdot s(\phi) , 
\end{equation}
where $s(\phi)$ is the window function which is define by

\begin{equation}
\label{eq:4}
\text{$s(\phi)$} = 
\begin{cases} 
\text{0, $\mid \phi-\varphi_p \mid > B$}  \\
\text{1, $\mid \phi-\varphi_p \mid \leqslant B$} ,
\end{cases}
\end{equation}
where $B$ is the selected angle range and the phase of $(\phi-\varphi_{p})$ is wrapped within $(-180^\circ,180^\circ)$. To include sufficient array elements and suppress the sidelobes, the angle range $B$ can be selected as 90$^\circ$. The modified classical beamforming result can be obtained by

\begin{equation}
\label{eq:5}
 Q(f,\phi) = \sum_{p=1}^{P} \omega_{p}(f,\phi)\cdot H_{p}(f)\ .
\end{equation}

It can also be expressed as a summation of the beam patterns of the $K$ paths by taking (\ref{eq:1}) into (\ref{eq:5}):
\begin{equation}
\label{eq:6}
\begin{split}
 Q(f,\phi) & = \sum_{k=1}^{K} \alpha_{k} \exp(-j2 \pi f \tau_{k})\ \sum_{p=1}^{P}  a_{p}(f,\phi_k) \cdot \omega_{p}(f, \phi)\  \\
 & =\sum_{k=1}^{K} \alpha_{k} \exp(-j2 \pi f \tau_{k}) \cdot \upsilon_k (f, \phi)\ ,
\end{split}
\end{equation}
where $\upsilon_k (f, \phi)$ represents the unit beam pattern of the $k$-th path with modified classical beamforming and $\mid \upsilon_k (f, \phi_k) \mid$ is the array gain. The PADP with modified classical beamforming can be obtained via inverse Fourier transformation:
\begin{equation}
\label{eq:7}
\begin{split}
 q(\tau,\phi) & = \sum_{f=f_L}^{f_U} Q(f,\phi)\cdot \exp(j2\pi f \tau)\ \\
 & = \sum_{k=1}^{K} \sum_{f=f_L}^{f_U} \upsilon_k (f, \phi) \cdot \alpha_{k} \exp[-j2 \pi f (\tau - \tau_{k})]  .
 \end{split}
\end{equation}
In the proposed method, $\upsilon_k (f, \phi)$ with a narrower beamwidth and lower side-lobes can be formed, which can mimic the Dirac delta function better and the multipaths can be identified as local peaks of the PADP. The peak identification is performed by firstly finding local maxima of power delay profile (PDP) curves \cite{haneda2015statistical}. Similar peak detection is also performed on angular domain at the identified delays. The estimated multipath is denoted by 
 \begin{figure}
\centering
{\includegraphics[width=0.5\columnwidth]{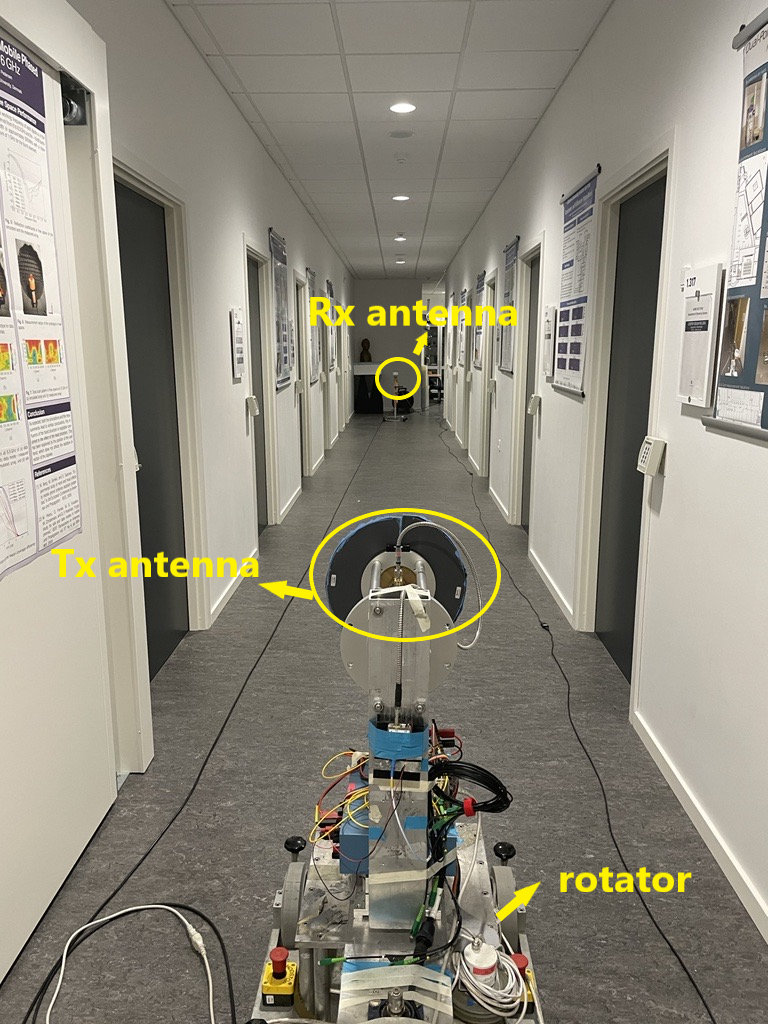}}

\caption{ A photo of the measurement scenario.}
\label{fig:scenario}
\end{figure}

\begin{figure}
\centering
{\includegraphics[width=\columnwidth]{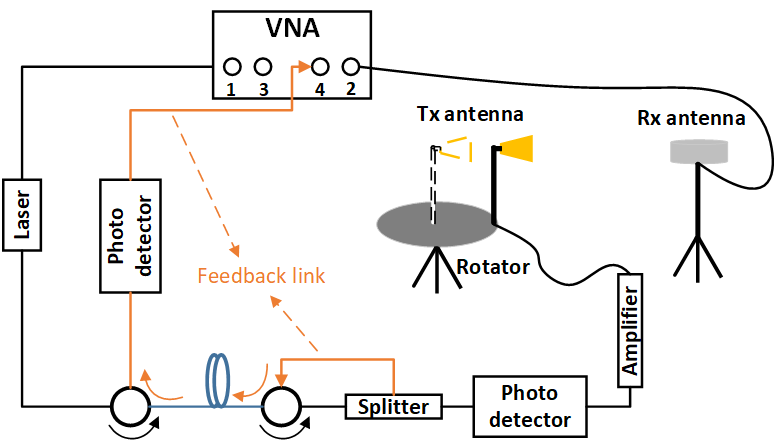}}

\caption{ The diagram of the measurement system.}
\label{fig:sounder}
\end{figure} 

\begin{figure*}
\centering
{\includegraphics[width=0.8\textwidth]{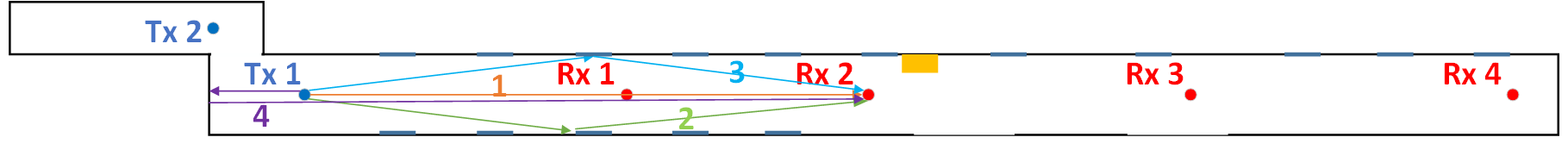}}

\caption{The diagram of the locations of Tx and Rx antennas in the measurements.}
\label{fig:locations}
\end{figure*} 

\begin{equation}
\label{eq:8}
\mathcal{P} = \{\phi_{k}, \tau_{k}, P_{k} \}_{k=1}^{K},
\end{equation}
where $P_{k}$ is the power of the $k$-th path. The estimated omni-directional pathloss can be calculated by 
\begin{equation}
\label{eq:9}
PL_{omni} = -10log_{10}(\sum_{k=1}^{k=K} \frac{P_k}{G_{tx} \cdot G_{rx} \cdot\mid \upsilon_k (f, \phi_k) \mid}),
\end{equation}
where $G_{tx} (\phi_k)$ and $G_{rx} (\phi_k)$ represent the gain of transmit (Tx) antenna and receive (Rx) antenna, respectively. Note that the array gain $\mid \upsilon_k (f, \phi_k) \mid$ at the center frequency can be used in (9) to calculate the pathloss when the gain of the directional antenna used for virtual array is approximately constant over the target frequency band.

\section{Measurement Campaigns}
\label{sec:campaign}

The measurements for both line-of-sight (LOS) and non-line-of-sight (N-LOS) scenarios were conducted in an indoor corridor at Aalborg University, as shown in Fig. \ref{fig:scenario}. Note that the measurement campaigns are designed only for validating the proposed method. Characterizing omni-directional pathloss for different deployment scenarios will be conidered in the future when massive measurement data are available. The measurement system is illustrated in Fig. \ref{fig:sounder} and its working principle is detailed in \cite{mbugua2019phase}. This VNA-based channel sounder can support long range phase-coherent measurements in the mm-Wave bands by using optical fiber cables and the phase compensation scheme. In VNA-based channel measurements, CFR is obtained from the measured S21 over the selected frequency band. 

Both the DSS measurements, i.e. with the directional antenna placed at the rotation center, and the proposed directional antenna based VAA measurements, i.e. with the directional antenna placed with a certain distance from the rotation center (as shown in Fig. \ref{fig:sounder}), were conducted. The DSS measurements and its result analysis are used for comparison. The Rx antenna is a vertically polarized biconical antenna \cite{biconical}, which has almost omni-directional radiation pattern in the azimuth plane (with around 2 dB ripple over 360 degrees) and a 5.5 dBi gain at 29 GHz. The corrugated antenna \cite{asy} with 40$^\circ$ HPBW and 13.5 dBi gain at 29 GHz is selected as the Tx antenna fixed on the mechanical rotator since it can form a VAA with high angular resolution as validated in \cite{li2022virtual}. The corrugated antenna was rotated automatically over the whole azimuth plane with a rotation step of 1.5$^\circ$ (i.e. 240 steps in total) for both the DSS and VAA measurements. The measured frequency range is from 28 to 30 GHz with 1001 frequency points for each rotation angle. For VAA measurements, the corrugated antenna was placed 0.15 m away from the rotation center to form a virtual UCA. As described, the measurement system and measurement time for the DSS and VAA schemes are the same.  

The locations of Tx and Rx antennas are shown in Fig. \ref{fig:locations}. For LOS scenarios, the Rx antenna was located at Rx 1 whereas the Tx antenna was located at Tx 1-4 to form direct links with various distances of 8 m, 14 m, 22 m, 30 m, respectively. For N-LOS scenarios, the Rx antenna was moved to Rx 2 where the LOS rays were blocked by walls and the locations of the Tx antenna remained the same as LOS scenarios with link distance of 11.4 m, 17.4 m, 25.4 m, 33.4 m, respectively. The Rx and Tx antennas were both placed over the ground with a height of 93 cm.

\section{Measurement Results}
\label{sec:results} 

\begin{figure}
\centering
{\includegraphics[width=\columnwidth]{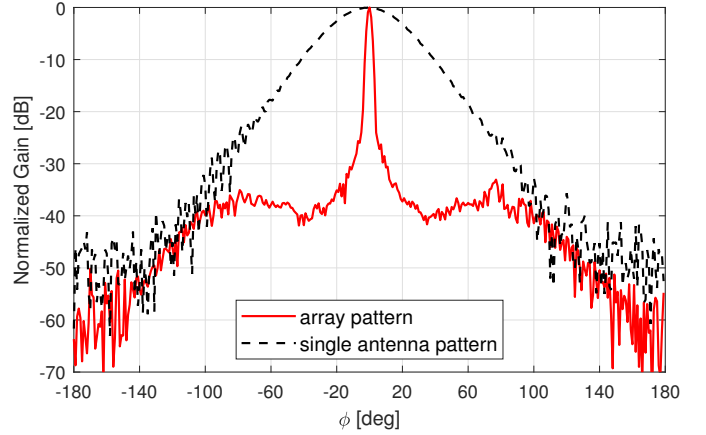}}

\caption{The measured corrugated antenna pattern (i.e. antenna pattern in the DSS scheme) and the simulated array beam pattern using VAA scheme at 29 GHz.}
\label{fig:pattern}
\end{figure} 

The measured radiation pattern of corrugated antenna and the simulated array beam pattern of the corrugated antenna based UCA at 29 GHz are compared in Fig. \ref{fig:pattern}. The UCA is composed of 240 elements with a radius of 0.15 m, which is the same as the settings in measurement campaigns. The window width $B$ is chosen as 90$^\circ$ according to \cite{li2022virtual}. It can be seen that the angular resolution of the corrugated antenna based UCA is greatly improved compared with the single antenna element. The high angular resolution and the low side-lobe level achieved by VAA can significantly improve the system's capability to identify  multipath in PADP.

\subsection{Power Angular Delay Profile}
\label{subsec:PADP}

\begin{figure}
\centering
{\includegraphics[width=\columnwidth]{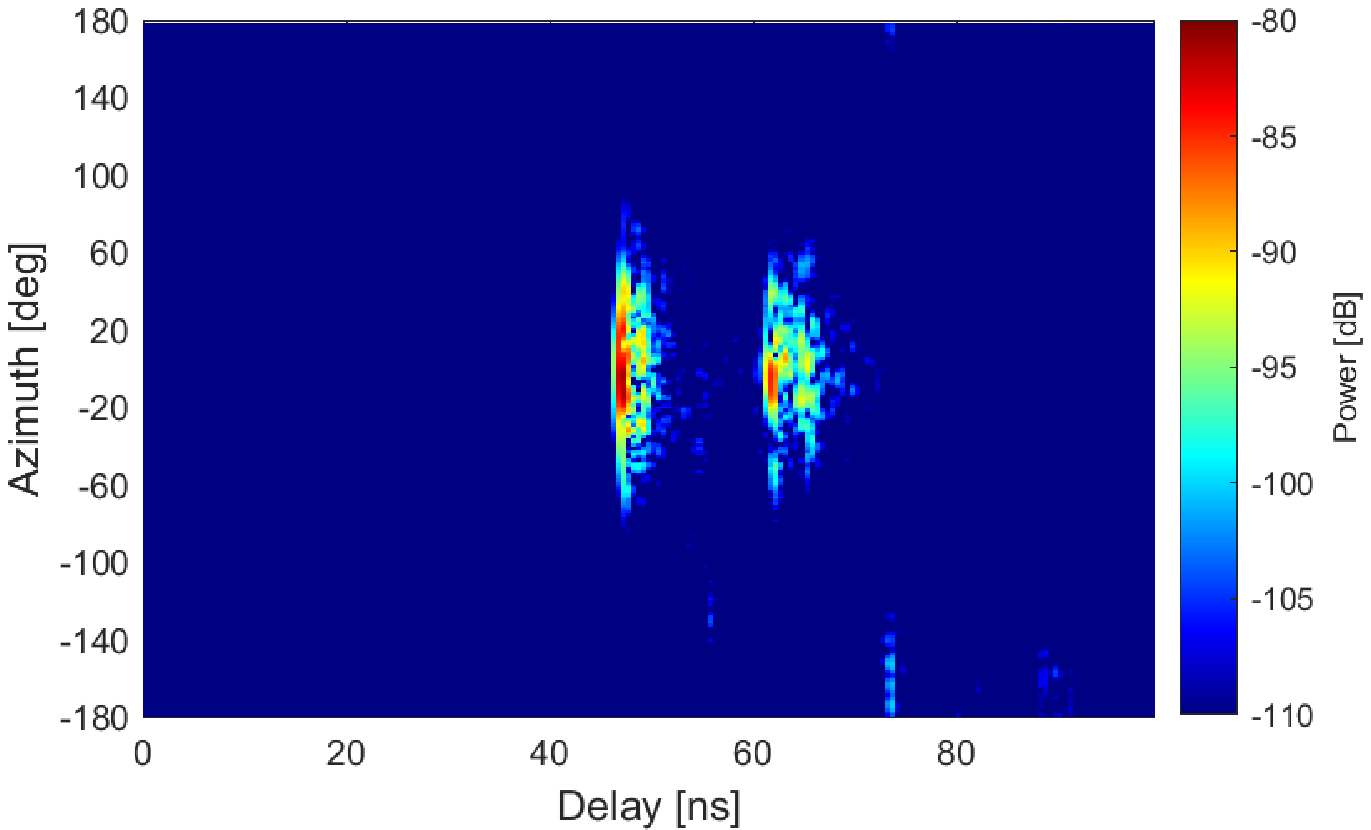}}

{\includegraphics[width=\columnwidth]{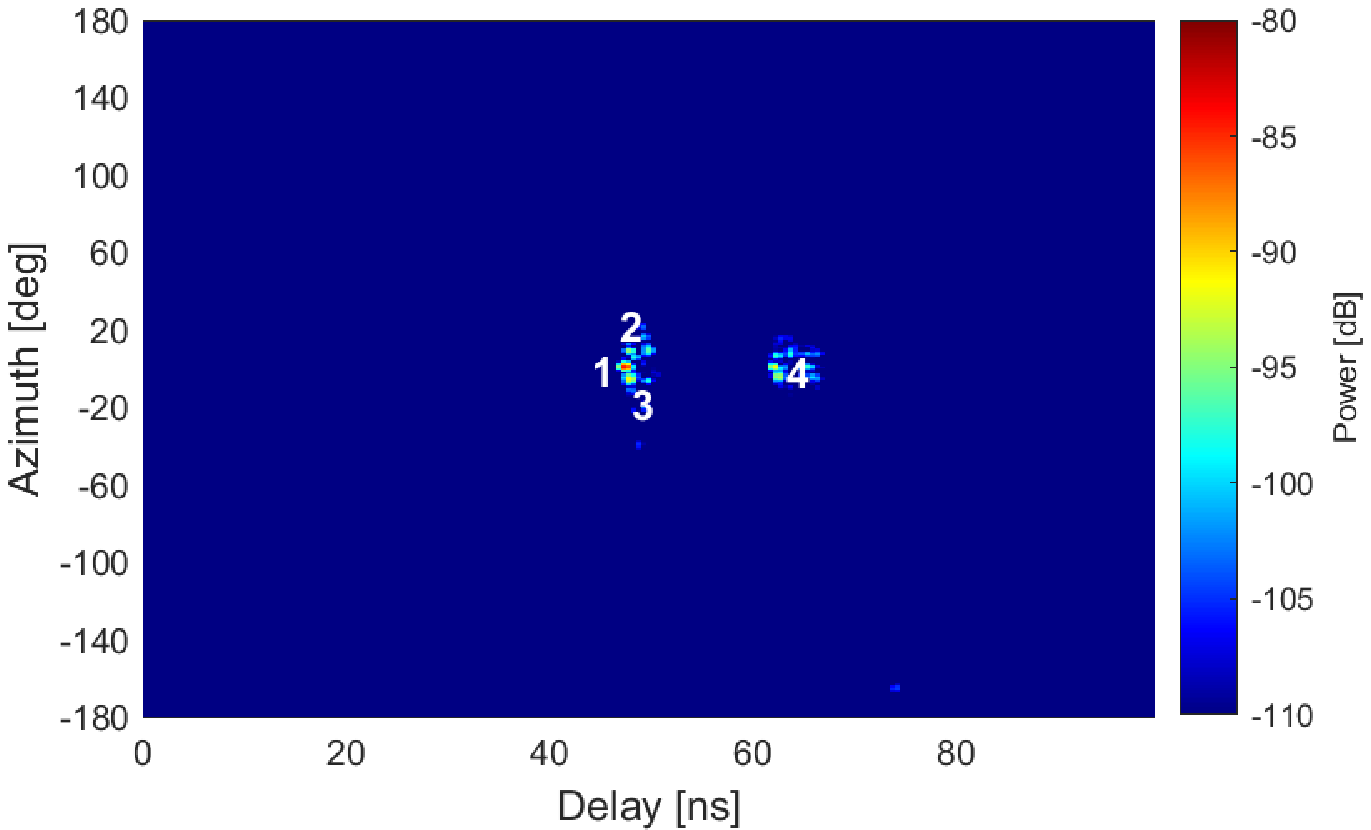}}

\caption{PADP obtained from (a) DSS and (b) VAA formed by corrugated antenna for LOS scenario with link distance of 14m.}
\label{fig:padp}
\end{figure}

\begin{figure}
\centering
{\includegraphics[width=\columnwidth]{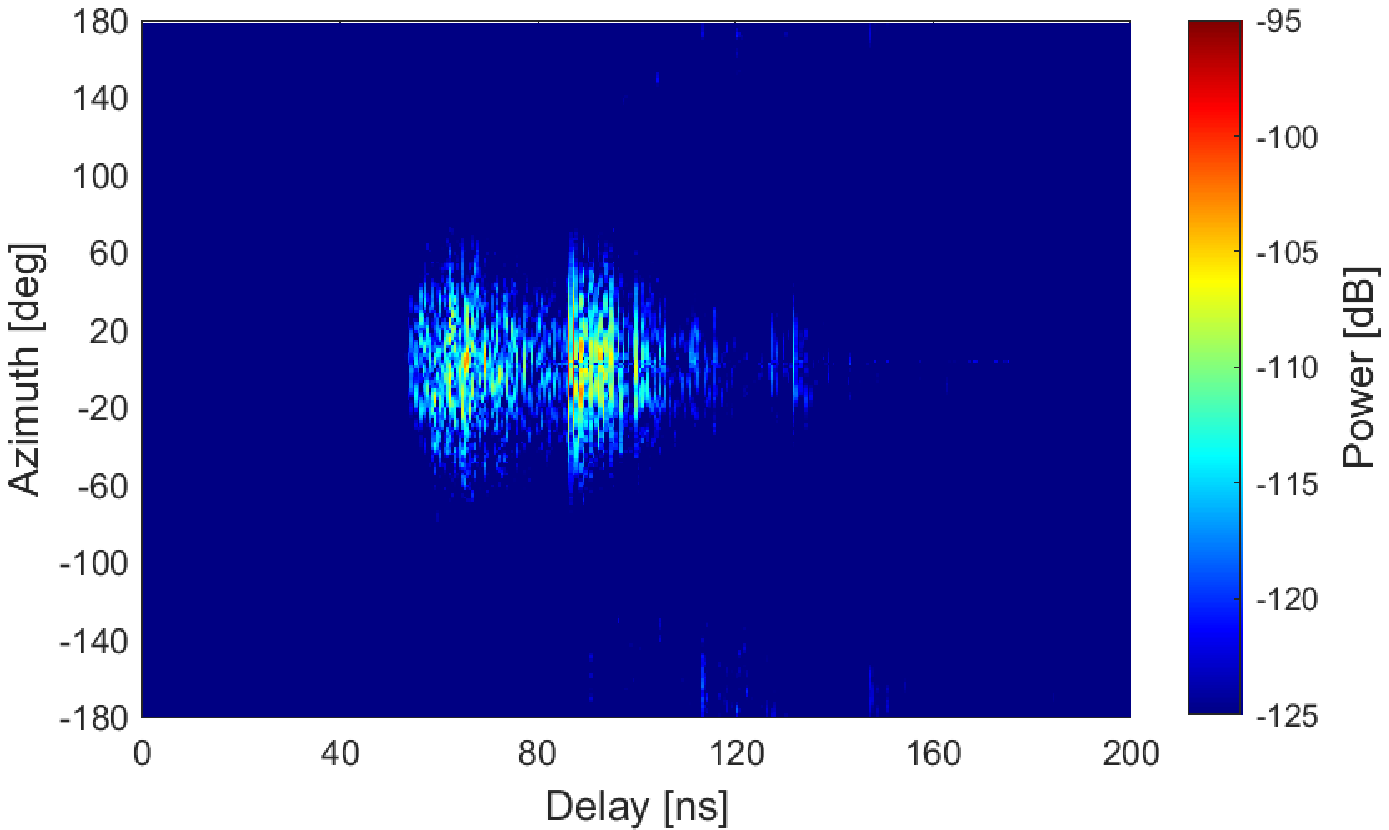}}

{\includegraphics[width=\columnwidth]{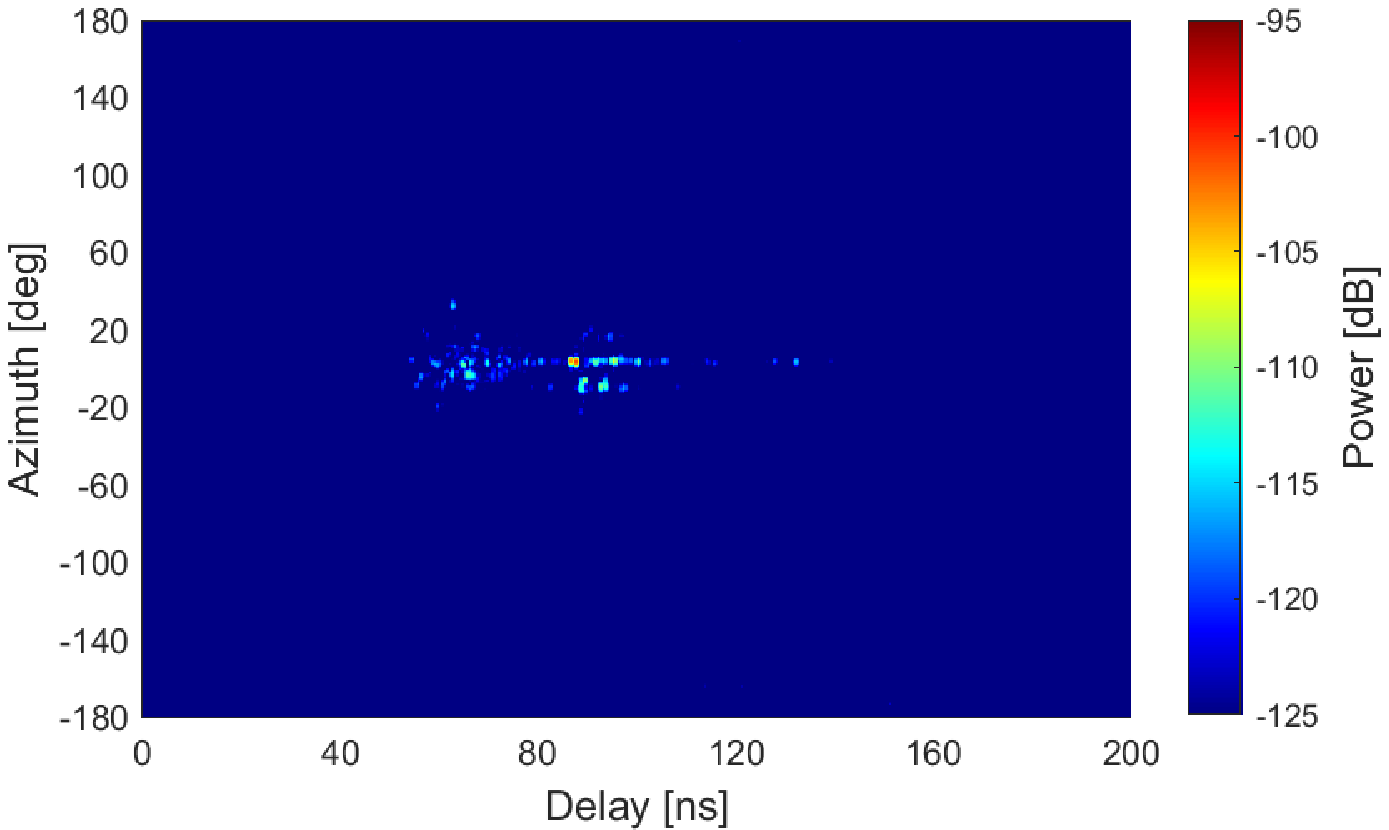}}

\caption{PADP obtained from (a) DSS and (b) VAA formed by corrugated antenna for N-LOS scenario with link distance of 17.4 m.}
\label{fig:padp2}
\end{figure} 

As mentioned in Section \ref{sec:campaign}, we have conducted 8 measurements in both LOS and N-LOS scenarios with different link distance. For simplicity, only the PADPs of LOS scenario with link distance of 14 m (Tx 1-Rx 2 shown in Fig. \ref{fig:locations}) and N-LOS scenario with link distance of 17.4 m (Tx 2-Rx 1 shown in Fig. \ref{fig:locations}) are given here as examples. The PADPs obtained from DSS and corrugated antenna based VAA scheme for LOS and N-LOS scenarios are depicted in Fig. \ref{fig:padp} and Fig. \ref{fig:padp2}, respectively. The distribution of the multipath in LOS scenario is relatively centralized whereas the multipath distributes in a wider angular and delay ranges in N-LOS scenario due to the high-order bounces caused by the walls. It can be clearly seen in Fig. \ref{fig:padp} (a) and \ref{fig:padp2} (a) that, different paths are overlapped with each other due to the wide main beam of the antenna.
 The omni-directional pathloss estimation can be distorted by the antenna pattern since the paths might be counted repeatedly at different rotation angles. Compared with the DSS method, the PADP obtained from VAA provides much higher angular resolution and the multipath becomes easier to be recognized. Four main propagation paths in the LOS scenario are marked in Fig. \ref{fig:padp} (b) and their trajectories in relation to the corridor geometry are shown in Fig. \ref{fig:locations}. The multipaths in N-LOS scenario are not traced back to the room geometry here since their trajectories are difficult to be determined with multiple high-order bounce paths.

\subsection{Omni-directional Pathloss}
\label{subsec:pathloss}

\begin{figure}
\centering
{\includegraphics[width=\columnwidth]{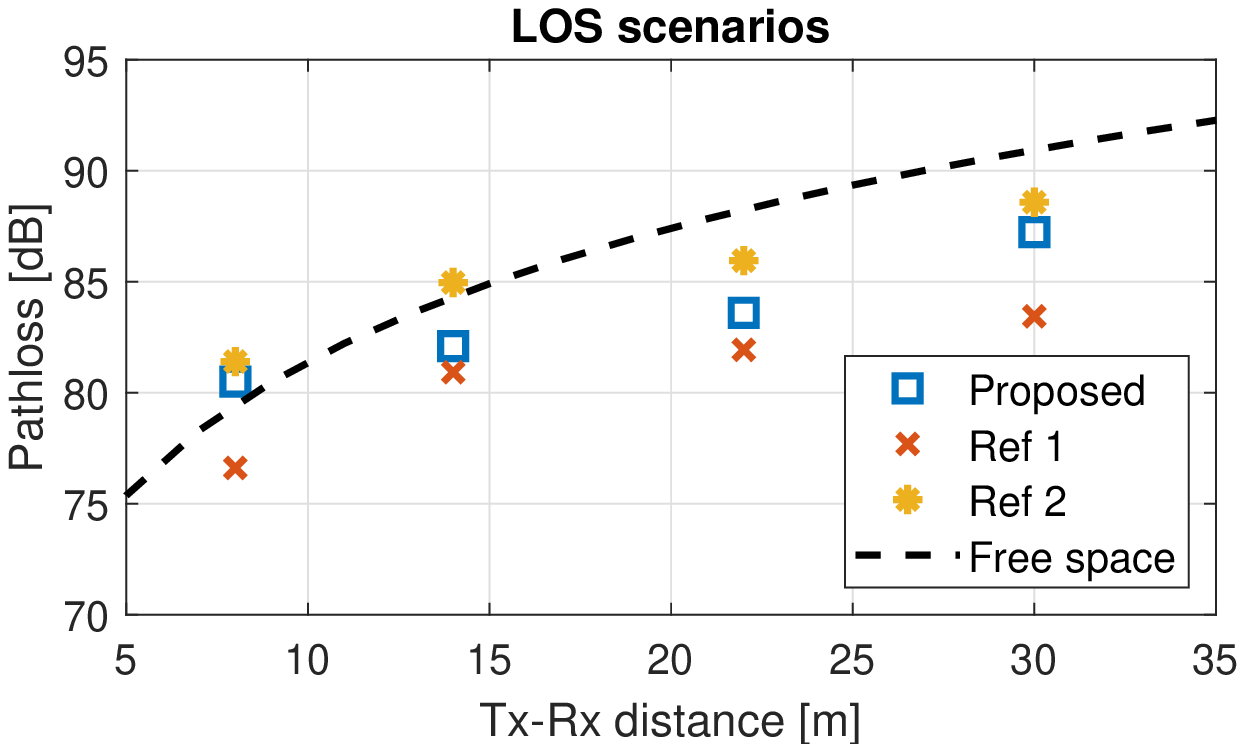}}

{\includegraphics[width=\columnwidth]{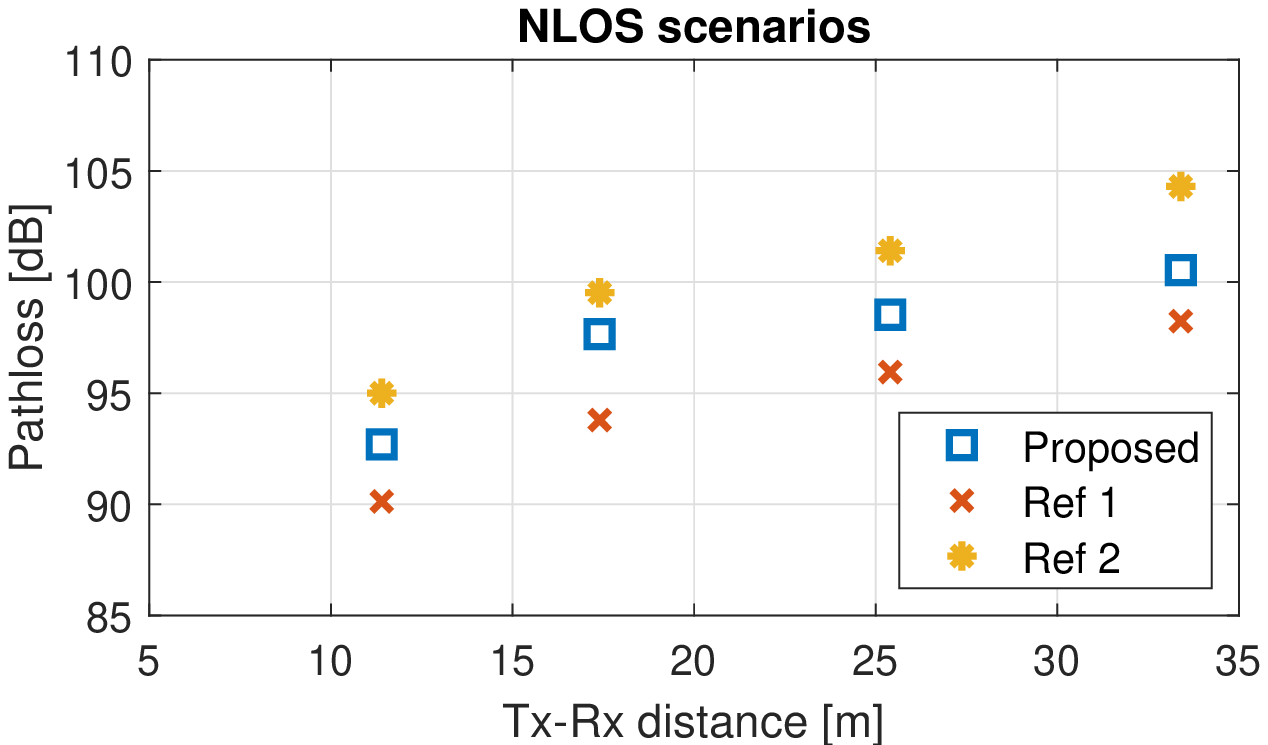}}

\caption{The estimated omni-directional pathloss at 29 GHz using conventional methods and the proposed method.}
\label{fig:pathloss}
\end{figure} 

Fig. \ref{fig:pathloss} depicts the estimated omni-directional pathloss using the proposed method based on VAA scheme and two reference methods mentioned in the introduction based on DSS scheme \cite{kyro2011measurement,akdeniz2014millimeter,sun2015synthesizing,maccartney2015indoor,fuschini2017analysis,hur2014synchronous,hur2015wideband}. \cite{kyro2011measurement,akdeniz2014millimeter,sun2015synthesizing,maccartney2015indoor,fuschini2017analysis} (denoted as Ref 1) employ the DSS and the pathloss is calculated by summing up all the powers in the PADP above the noise level. \cite{hur2014synchronous,hur2015wideband} (denoted as Ref 2) also use the DSS and only the powers of the identified propagation paths are considered in omni-directional pathloss estimation. Free space pathloss is also given in Fig. \ref{fig:pathloss} as a reference. As discussed earlier, the proposed method uses corrugated antenna based VAA together with the modified classical beamforming to obtain high angular resolution in the PADP. Therefore, the multipath can be more accurately detected by finding local maximum peaks in both angular and delay bins. The results of the proposed method and Ref 2 in Fig. \ref{fig:pathloss} (a) are comparable to the free space loss whereas the pathloss obtained from Ref 1 has higher deviation. The estimated pathloss obtained from Ref 1 is always below the proposed method with a maximum value of 3.9 dB in both LOS and N-LOS scenarios. The reason is that the same paths are counted repeatedly at different rotation angles. The estimation difference between Ref 2 and the proposed method is up to 2.8 dB in LOS scenarios and reaches 4 dB in N-LOS scenario. The pathloss is overestimated using Ref 2, especially in N-LOS scenarios since its capability of identifying multipath is constrained by the non-ideal antenna pattern. Only the path with maximum power is counted at the identified delay bins in this case. Many weak components, which have similar delays with a strong path cannot be detected. Overall, the pathloss calculated from the proposed method is always between the pathloss obtained from two reference methods, as expected.

\section{Conclusion}
\label{sec:conclusion}
A new approach of estimating the omni-directional pathloss is presented in this letter. Directional antenna based virtual array is used to achieve high angular resolution and reduce the side-lobe effects on the PADP. The multipath can be identified by searching local maxima in the power-angle-delay spectra and the omni-directional pathloss is then calculated based on the powers of the detected paths. Compared with the traditional DSS method, the propagation paths are more accurately identified using VAA scheme without introducing additional cost and increasing the measurement time. The measurement results show that the estimation difference between the proposed method and Ref 1 is up to 3.9 dB in both LOS and N-LOS scenarios. Compared with the results obtained from Ref 2, the difference can reach 2.8 dB and 4 dB in LOS and N-LOS scenarios, respectively. As observed, the estimation errors introduced by non-ideal antenna patterns using DSS method are not negligible. The proposed method can provide a more accurate path detection and improve the omni-directional pathloss estimation results.


\bibliographystyle{IEEEtran}
\addcontentsline{toc}{section}{\refname}\bibliography{Li_library}

\end{document}